\documentclass[11pt]{article}
\usepackage{moriond,epsfig}
\usepackage{amsmath,amssymb,amsbsy,epsfig}
\newcommand{\lum}[1]{$10^{#1}$ cm$^{-2}$s$^{-1}$ }

\def\Journal#1#2#3#4{{#1} {\bf #2}, #3 (#4)}

\def\PRD{{\em Phys. Rev.} D}

\def\be{\begin{equation}}
\def\ee{\end{equation}}
\def\bea{\begin{eqnarray}}
\def\eea{\end{eqnarray}}

\begin{document}
\vspace*{4cm}
%\title{PHYSICS WITH CMS AND AN INTEGRATED LUMINOSITY OF 10 fb$^{-1}$}
\title{BRIEF PHYSICS SURVEY WITH CMS IN YEAR ONE}

\author{ G. BRUNO }

\address{CERN, EP Division, CH-1211 Geneve 23, \\
Geneva, Switzerland}

\maketitle\abstracts{
The CMS detector is one of the two general purpose experiments that will 
study the collisions produced by the Large Hadron Collider (LHC). The LHC 
is supposed to start its operation in 2007 at an instantaneous luminosity of
$2\times$\lum{33}, which may well result in an integrated luminosity of 
10 fb$^{-1}$ after the first year of running. 
The corresponding physics reach of CMS is exemplified with the study
of a few standard model channels (weak boson and top quark
production) and with the searches for Higgs bosons.
}

\section{Introduction}
The Large Hadron Collider (LHC) is designed to produce pp
collisions at a nominal 
luminosity of \lum{34}. With the foreseen startup luminosity
of $2\times10^{33}$ cm$^{-2}$s$^{-1}$, 
the LHC is expected to deliver an integrated luminosity of 
about 10 fb$^{-1}$ in the first year of running. 
%A general purpose experiment
%running at the LHC can largely benefit from such an amount of physics input.
The corresponding physics reach of the CMS experiment~\cite{cmstp} is briefly
surveyed in this document, both with Standard Model (SM) studies
(Section 2) and searches for Higgs bosons within the SM
(Section 3) and its minimal supersymmetric extension, the MSSM (Section 4).
%The physics reach of the CMS experiment~\cite{cmstp} 
%in the Standard Model (SM) and in the Minimal
%Supersymmetric Standard Model (MSSM) Higgs sector is reviewed in
%this document. 
%Expected yields of W and Z
%bosons as well as top quark are also given.
The PYTHIA Monte Carlo generator~\cite{ref8} was normally used to
generate the primary interactions for all production
processes. The detector simulation
and reconstruction packages CMSIM~\cite{ref9} and ORCA~\cite{orca}
were used to accurately
reproduce the response of the detector for benchmark
channels. However, to allow large event samples to be produced and
statistically robust studies to be performed
the fast simulation package CMSJET~\cite{cmsjet}, parameterised
to reproduce the results of the complete simulation for
particle identification, {\rm b}-tagging, missing
energy resolution and jet reconstruction, was often preferred. In both cases
the online selection chain, which consists of the Level-1
trigger and the High-Level Trigger (HLT), was simulated~\cite{daqtdr}.

The physics reach after the first year of running is closely
related to the level of understanding of the detector behaviour. 
In particular, the accuracy that will be achieved in
the alignment of the inner tracker and muon detectors and the
calibration of the calorimeters is expected to play a crucial role. 
All the results presented in this document assume a perfectly aligned
and calibrated detector.

%For {\em in situ} alignment and calibration an important role
%is played by the leptonic (electrons and muons) decay channels of the W and
%Z bosons.

%All results have been obtained in either full or fast simulation studies.
%In the former case the particles resulting from simulated collisions 
%are simulated with PYTHIA~\cite{ref8} and the produced particles 
%are propagated through the apparatus by means of a detailed
%GEANT3-based simulation of the
%CMS detector~\cite{ref9}. Then, the response 
%of the detectors is accurately simulated~\cite{orca} and, finally, the
%reconstruction algorithms developed for CMS~\cite{orca}
%are run to obtain the physics observables.
%On the other hand, fast simulation will mean throughout this document a
%study in which the steps going from 
%particle propagation to basic reconstruction are replaced by the use
%of parameterization functions
%implementing selection and reconstruction efficiency as well as
%resolution for the physics objects relevant to 
%the study. Fast simulations are unavoidable whenever the statistics
%needed is so high that it would require unacceptably long computing time.

\section{Standard model physics}
Standard model particles, particularly {\rm W} and {\rm Z} bosons and
{\rm t} quarks, will be copiously produced at the LHC, and
easily detectable at startup through their leptonic decays
(electrons and muons). At a centre-of-mass energy of
%The measurement of W and Z boson production properties, and especially their
%couplings, will be one of the  fundamental topics to be studied
%at the LHC. On one hand, deviations may hint at new physics. On the other,
%first manifestations of supersymmetry
%may have to be discriminated against a background of $W$+jets and $Z$+jets
%events. The same holds for heavy flavour physics and, in particular, 
%studies of the $t$ quark, where couplings, rare decay modes, spin 
%characteristics and correlations have to be studied. $W$, $Z$ and $t$ quark 
%production also provide key tests of QCD.\\ 
%The main channels for analysis of the $W$, $Z$ and $t$
%quark at the LHC will be their decays with leptons in the final state.
14 TeV in pp collisions 
the cross sections in final states with leptons, ${\rm W} \to
\ell\nu_\ell$, ${\rm Z} \to \ell^+\ell^-$ and ${\rm t\bar t} \to
{\rm W}^+{\rm b} {\rm W}^-\bar{\rm b} \to \ell\nu_\ell + {\rm
  X}$, amount to approximately 20, 2 and 0.13 nb per lepton
generation, respectively,
which correspond to event production rates of 40, 4 and 0.26 Hz at
startup luminosity. Table 1 shows the corresponding event yields of
the muon channels expected with
$10\,{\rm fb}^{-1}$ in the detector acceptance
when the Level-1 trigger and the HLT
transverse momentum thresholds are applied (17 and
9\,GeV/$c$ at the HLT for single and dimuon events, with at least
one muon isolated).
%
%
%the channels ${\rm W} \to \mu \nu$,
%${\rm Z}\to \mu \mu $ and ${\rm t \bar{t}} \to {\rm Wb Wb} \to \mu \nu
%+ {\rm X}$ are approximately
%20 nb, 2 nb and 0.13 nb respectively. These numbers correspond
%to event production rates of 40 and 4 Hz for the {\rm W} and {\rm Z} bosons respectively. 
%The CMS muon trigger system covers the pseudorapidity region
%$\lvert\eta\rvert <2.1$ and, since the magnetic field configuration is,
%with good approximation,
%such that particles do not change their $\eta$
%along their path, one can reasonably define the acceptance as the
%percentage of events in which there is at least a muon with
%$\lvert\eta\rvert <2.1$ at generation.  
%The full selection chain, consisting of the
%L1 trigger and of the High Level Trigger (HLT) algorithms, has been
%simulated~\cite{daqtdr} on several
%signal samples produced with Pythia~\cite{ref8}.
%It has been demonstrated that the on-line selection
%can be accomplished  at start-up by setting the HLT transverse
%momentum threshold on single
%isolated muon and di-muon events (one of them isolated) at 19 and 7
%GeV/$c$ respectively,
%the corresponding L1 thresholds being softer. 
%The selection efficiency 
%for the three channels mentioned above (considering only the events in 
%the geometric acceptance) is presented in  Tab. 1.
%In the same table the yields expected with an integrated luminosity
%of 10 fb$^{-1}$ are also reported. The
%yield is calculated assuming the cross section, acceptance and
%efficiency values that are reported in the same table. 
  \begin{table}[htb]
    \label{yield}

    \begin{center}
      \begin{tabular}{|c|c|c|c|c|} 
        \hline
        & $\sigma$ (nb) & Acc. (\%)  & Eff. (\%)
        & Yield for 10 fb$^{-1}$   \\ 
        \hline 
        ${\rm W}\to \mu \nu_\mu$ & 19.6 & 50 & 69 & $7\times 10^7$ \\ 
        \hline
        ${\rm Z}\to \mu^+ \mu^-$ & 1.84 & 71 & 92 & $1.1 \times 10^7$ \\ 
        \hline
        $t\bar{t}\to {\rm W^+bW^- \bar b} \to \mu \nu_\mu + {\rm X}$ & 0.126 & 86 & 72 & $7.8 \times 10^5$ \\ 
        \hline
       \end{tabular}
    \end{center}
   \caption{Cross section, geometric acceptance, online selection efficiency
   and yield for some representative SM decay modes with at least one muon in
   the final state. The geometric acceptance is defined as the
   percentage of events with at least one muon at generation  
   in the muon system nominal coverage: $\lvert\eta\rvert <2.1$. The
   Level-1 and the HLT  selection is tuned for the initial luminosity of
   $2\times10^{33}$ cm$^{-2}$s$^{-1}$ expected at the LHC.}
\end{table}
In one year, between ten and hundred million {\rm Z} and {\rm W} events
are therefore expected with at least one muon in the detector acceptance. These
events (and those with electrons in the final state) are foreseen to
be primarily used for tracker and muon
system alignment and calorimeter calibration purposes. 
A survey of the methods foreseen for
each sub-detector can be found in~\cite{daqtdr} and in the references
therein.
The {\rm Z} events will be used to set the lepton
energy and momentum scale, which is expected to be one of the major sources of
uncertainty on several precise measurements of the SM parameters 
like the {\rm W} mass. 
It is foreseen that with the first 10 fb$^{-1}$ the
error on the {\rm W} mass will be already of the order of 30\,MeV~\cite{lhcwork}, 
which is better than what achieved at LEP~\cite{h_upperlim} and
comparable to the
ultimate expectations of the Run II at the Tevatron~\cite{teva}.
A full review on the SM physics to be done at the LHC
can be found in~\cite{lhcwork}.
%Possible hints at new physics can be inferred if these precision measurements
%deviate from the SM expectation.
%The W and Z events, as well as the million top quark pairs,
%also provide key tests of QCD [examples + refs, please].
Finally, a detailed understanding of the {\rm W}, {\rm Z} and {\rm t} quark
pairs events is of importance in
the quest for new physics as they often constitute the major
background sources to, {\it e.g.}, Higgs boson or Supersymmetry
searches.

\section{The search for the SM Higgs boson}
\begin{figure}[hbtp]
  \begin{center}
   \mbox{\epsfig{file=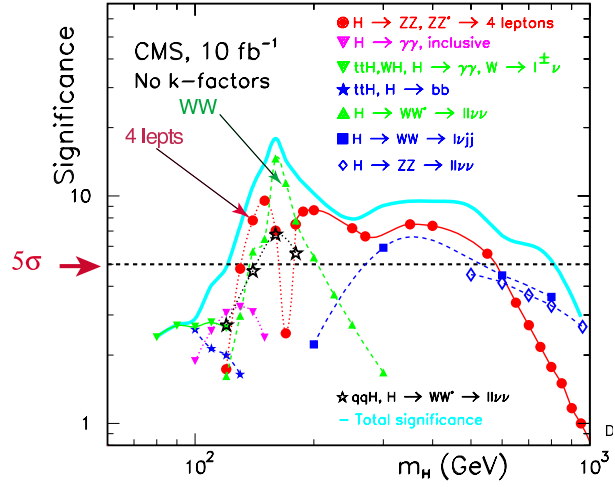, clip,width=8cm}}
%   \mbox{\epsfig{file=h_gamgam.eps, clip,width=5 cm}}
  \caption{Expected statistical significance ($S/\sqrt{B}$) with 10
  fb$^{-1}$ for the SM Higgs boson as a function of $m_{\rm H}$. Leading
  order cross sections are used for all processes.}
 \label{h_discrange1}
  \end{center}
\end{figure}

\begin{figure}[hbtp]
  \begin{center}
   \mbox{\epsfig{file=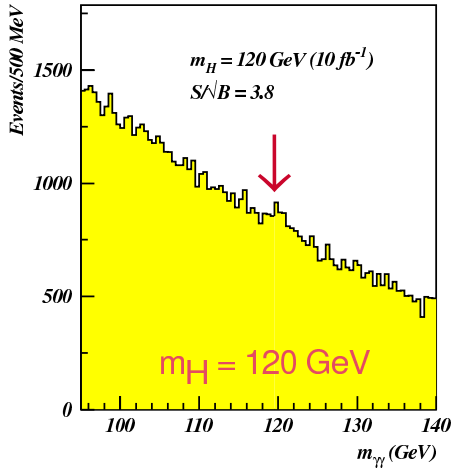, clip,width=6 cm}}
   \caption{${\rm H}\to \gamma \gamma$ ($m_{\rm H}=120$ GeV/$c^2$) signal on the top of the background with 10 fb$^{-1}$.}
    \label{h_gamgammass2} 
  \end{center}
\end{figure}

All existing direct searches and precision
measurements performed at LEP and SLD are compatible with the existence
of a SM-like Higgs boson of mass between 114.4~\cite{h_lowerlim}
and 211\,GeV/$c^2$~\cite{h_upperlim} at 95\% C.L. A heavier Higgs boson,
however, can be consistent with the precision electroweak measurements
in models more
general than the minimal standard model~\cite{peskin}. The CMS experiment was
designed to extend the mass range in which a discovery could
take place all the way up to about 1\,TeV/$c^2$.
%The search for the Higgs boson is one of the most important tasks of
%the general purpose experiments at the LHC. The mass range where discovery
%could take place starts at the present upper limit of 114.4 GeV/$c^2$
%set by LEP~\cite{h_lowerlim} and spans up to about 1 TeV/$c^2$. 
The expected statistical significance of such a discovery
with an integrated luminosity of 10 fb$^{-1}$ is displayed in
Fig.~\ref{h_discrange1}~\cite{h_survey} for the most
relevant signals
observable  with the CMS detector and for their combination as a
function of the Higgs boson mass ($m_{\rm H}$). 
%The left plot in Fig.~\ref{h_discrange1}~\cite{h_survey}
%shows the expected statistical significance of the most relevant
%SM Higgs signals observable with the CMS detector as a function of the
%Higgs mass ($m_{\rm H}$) and for an integrated luminosity of      
%10 fb$^{-1}$. 
In this figure, the
signal and background cross sections were computed at leading order
only (the inclusion of k factors would change the expected
significance by $k_{\rm signal}/\sqrt{k_{\rm background}}$).
%No k-factors have been used in this plot
%neither for the signals nor for the backgrounds. 
\begin{itemize}
\item As can be seen in Fig.~\ref{h_discrange1}, the spectacular signature
from the Higgs boson decay into four charged leptons, 
${\rm H}\to {\rm ZZ} \to \ell^+\ell^- \ell^+\ell^-$, allows the
mass range from 130 to 600\,GeV/$c^2$ to be covered~\cite{h_4eref}, with a
gap around the {\rm WW} threshold.
\item To cover the gap caused by the drop of the {\rm ZZ}
branching fraction in this region, the decay 
${\rm H} \to {\rm W}^+{\rm W}^- \to \ell^+\nu_\ell\ell^-\bar{\nu_\ell}$
is best suited for an early discovery~\cite{h_lnulnuref}. The most
important sources of background
to this channel are ${\rm t\bar{t}}$,  {\rm WW} and {\rm Wt}
production. It has been demonstrated that 
jet veto in the central rapidity region allows the ${\rm t\bar{t}}$
and {\rm Wt} backgrounds to be suppressed while {\rm WW} spin correlation
effects are an effective tool against the {\rm WW} continuum production.
\item Finally, Higgs boson masses below 130\,GeV/$c^2$ need
the contribution from the $\gamma\gamma$, ${\rm b
\bar b}$ and $\tau^+ \tau^-$ (not shown in Fig.~\ref{h_discrange1})
decay channels, in addition to all of the above.
These decay channels are expected to be particularly difficult because
of large backgrounds. The difficulty of the task is illustrated in
Fig.~\ref{h_gamgammass2}, 
which shows the $\gamma\gamma$ invariant mass spectrum in
the presence of a SM Higgs boson with mass
120\,GeV/$c^2$~\cite{h_gamgamref}. 
Next to leading order cross sections are used for
both signal and background. Events are selected requiring in the HLT
two isolated photons having transverse momenta greater than 25 and 40 GeV/$c$.
%which correspond to the cuts to be applied in the HLT selection.
Clearly, this channel relies on an excellent
electromagnetic calorimeter calibration.
% and  a discovery in this mass
%region within a year requires a fully aligned and calibrated
%detector.
\end{itemize}
%Next to leading order cross sections are used for both signal and
%background. Events are selected requiring two isolated photons having
%transverse momenta greater than 25 and 40 GeV/$c$, which correspond
%to the cuts to be applied in the HLT selection. The signal
%significance is 3.8. This result demonstrates that
%an integrated luminosity of 10 fb$^{-1}$ is not enough for discovery.\\
%especially in view of the fact that the CMS electromagnetic
%calorimeter will not have the expected energy resolution in the first
%year of running.
Despite the lower cross section with respect to the gluon fusion
production mechanism (${\rm gg} \to {\rm H}$), 
signal events produced in the weak
boson fusion channel (${\rm qq \to qq H}$) display the additional feature of
two energetic jets in the forward and backward directions.
In addition, the absence of colour exchange in the hard process leads
to a low jet activity in the central rapidity region. These two
features allow high rejection of the ${\rm t \bar t}$, single {\rm W} and
Z production accompanied by jets and QCD multi jet backgrounds. 
Very interesting results have been obtained for the decay
channels of a low mass Higgs boson produced via the weak
boson fusion process: $\gamma
\gamma$~\cite{h_gamgam}, ${\rm b \bar b}$~\cite{hbb}, $\tau^+
\tau^-$~\cite{htautau} and also ${\rm W}^+{\rm W}^- \to
\ell^+\nu_\ell\ell^- \bar{\nu_\ell}$~\cite{h_wworca}.

%Simulation studies where constraints are imposed
%on the presence of such jets are ongoing~\cite{h_wworca}~\cite{h_gamgam}.

\section{Higgs boson searches in the MSSM}
\begin{figure}[hbtp]
  \begin{center}
   \mbox{\epsfig{file=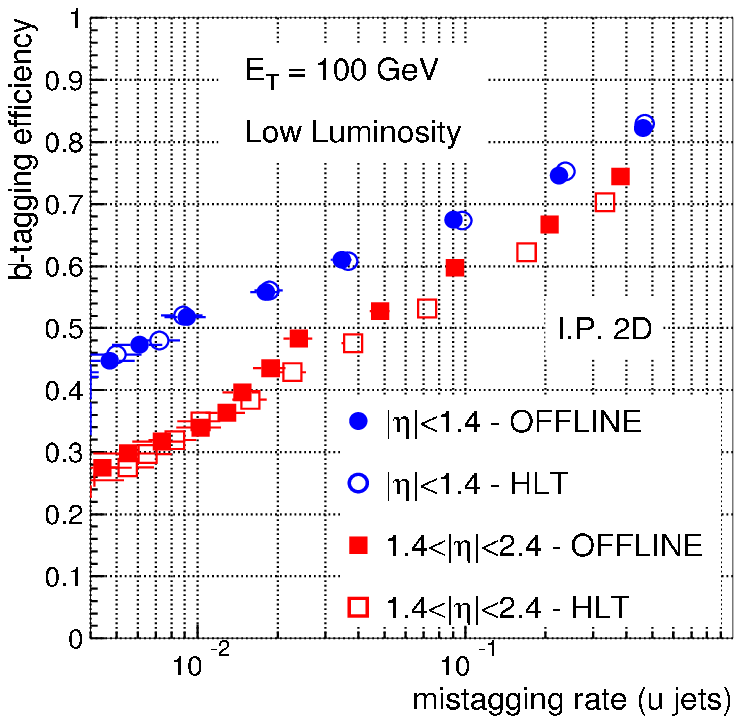, clip,width=7cm}} 
%\hspace{2 cm}
%   \mbox{\epsfig{file=a_tautau1.eps, clip,width=5cm}}
   \mbox{\epsfig{file=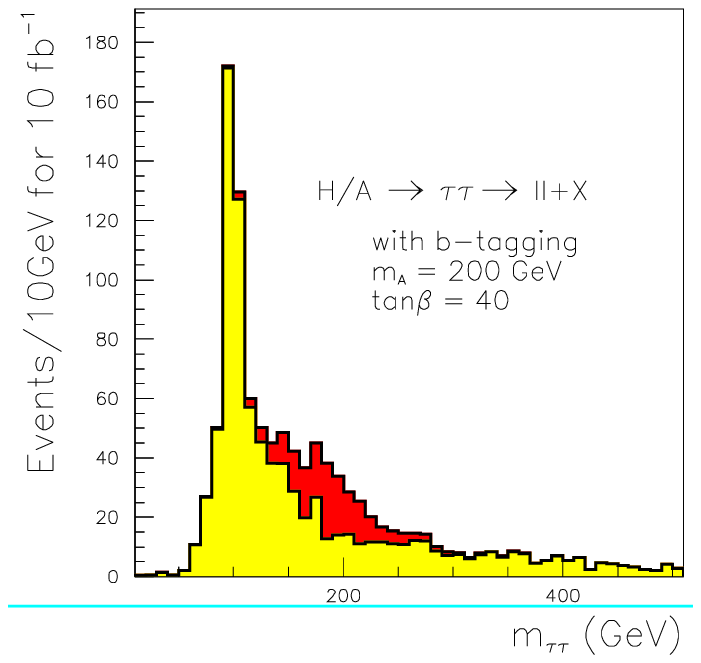, clip,width=7cm}} 
    \caption{Left: Efficiency of the $b$-tagging versus mistagging rate for
  jets with transverse energy of 100 GeV. Right: Reconstructed mass distribution of the ${\rm H/A} \to \tau^+ \tau^- \to \ell^+\nu_\ell\ell^-\bar\nu_\ell$ signal and background with 10 fb$^{-1}$.}
    \label{atau2}
  \end{center}
\end{figure}

\begin{figure}[hbtp]
 \begin{center}
   \mbox{\epsfig{file=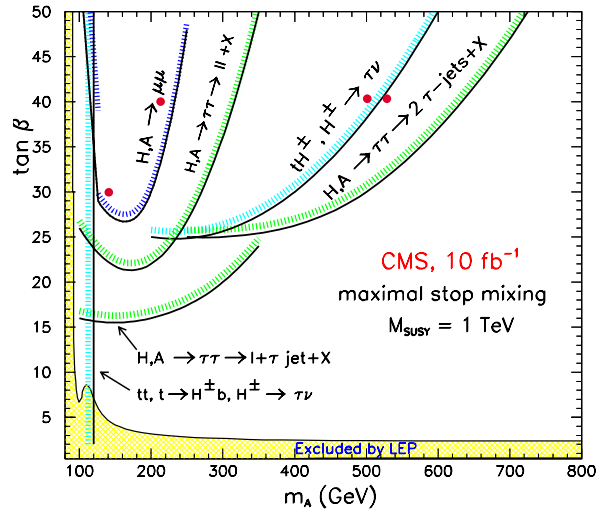, clip,width=8cm}} 
   \caption{Expected $5\sigma$ discovery reach with 10 fb$^{-1}$ for the
  MSSM heavy Higgs bosons in  CMS in the maximal mixing scenario  as a
  function of $m_{\rm A}$ and $\tan{\beta}$.}
    \label{h_discrange2}
  \end{center}
\end{figure}

%The MSSM~\cite{mssm} is one of the extensions of the SM that would permit 
%to eliminate some theoretical SM drawbacks. Supersymmetric 
%theories are also attractive as they provide a theoretical
%framework in which to include all known fundamental interactions.
In the MSSM, at least two Higgs doublets are needed,
in contrast to the minimal standard model in which only one
Higgs doublet is necessary. After three degrees of freedom
are used to give masses to the {\rm W} and {\rm Z} bosons, five physical states
remain: two charged Higgs bosons ${\rm H}^\pm$ and three neutral
Higgs bosons h, H and A (h and H are scalars while A is
pseudoscalar).
%A striking prediction of the MSSM is that the lighter scalar Higgs  has
%to have a mass smaller than 130\,GeV/$c^2$.
The lighter scalar Higgs boson {\rm h} is expected to have a mass smaller than
130\,GeV/$c^2$ and to behave like the SM Higgs
boson over most of the $\tan{\beta}$-$m_{\rm A}$ parameter space,
where $\tan{\beta}$ is the ratio of the vacuum expectation values of the two
Higgs doublets.
Its discovery, therefore, needs the full statistics of the first year
and a well aligned and calibrated detector.
The heavier neutral Higgs bosons {\rm H} and {\rm A}, however, benefit from an
enhancement of their coupling to {\rm b} quarks and $\tau$'s by $\tan{\beta}$.
%As a consequence, the production of the {\rm A} and {\rm H} Higgs bosons
%through the ${\rm gg}$, ${\rm q\bar{q} }\to {\rm b
%\bar b H/A}$ processes
%associated production with a pair of $b$ quarks 
%may be strongly enhanced with respect to the SM: 
%for $\tan{\beta} \gtrsim 10$ and $m_{\rm A}\gtrsim 300$ GeV/$c^2$
%these processes contribute more than 90\% of the total Higgs boson
%cross section. 
Furthermore,
the large branching fraction at high $\tan{\beta}$ of
${\rm A/H} \to \tau^+ \tau^-$ and ${\rm H^{\pm}} \to
\tau \nu_\tau$ makes these final states best suited
for discovery over a significant portion of the
$\tan{\beta}$-$m_{\rm A}$ parameter space.
%, as displayed in Fig.~\ref{atau1}.
Therefore, good {\rm b}- and $\tau$-tagging capabilities, to be
performed also online, are crucial for this searches. 
Several $\tau$- and {\rm b}-tagging algorithms have been developed in CMS, as
reported in~\cite{daqtdr} and in the references therein. 
The algorithms that have been tested in the HLT selection use the
information from the calorimeters,
which define cones of interest centred around the
reconstructed jets, and from the innermost layers of the tracker
detector, which allow the primary vertex and the
parameters of the tracks belonging to the jets to be determined. 
%In some cases track
%reconstruction is refined with the information from a few layers of
%the silicon strip detectors. 
The identification of a $\tau$ jet is
based on the reconstruction of an isolated collimated jet. 
%and uses information from the calorimeters and pixel
%detectors. Optionally, track reconstruction can be refined with the
%information from a few layers of
%the silicon strip detectors. The same information is used in ${\rm b}$
%tag algorithms  
The left plot in Fig.~\ref{atau2} shows the performance of a {\rm b}-tagging
algorithm~\cite{daqtdr} that relies on the measurement of the 
track impact parameter.
The version of the algorithm, indicated as ``HLT'' in
the plot, has proved to be sufficiently robust and fast for use
in the CMS HLT selection. 
The reconstructed $\tau^+\tau^-$ invariant mass distribution~\cite{a_tautauref}
for $m_{\rm A} = 200$\,GeV/$c^2$ and
$\tan\beta = 40$ in the fully leptonic final state of the heavy
neutral Higgs bosons ($\tau^+\tau^- \to
\ell^+\nu_\ell\ell^-\bar\nu_\ell$) after background
rejection with {\rm b}-tagging is shown in the right plot of Fig.~\ref{atau2}.
%Results from one of such studies~\cite{a_tautauref} are shown
%in Fig.~\ref{atau2} where
%the reconstructed mass distribution of the heavy neutral Higgses
%in the $\tau \tau \to ll \nu \nu$ final states is reported. 
A significant excess over the background is clearly visible with 10 fb$^{-1}$. 
Hadronic $\tau$ decays increase the search sensitivity to larger
values of $m_{\rm A}$ as can be seen in 
Fig.~\ref{h_discrange2}, which shows the coverage of the parameter space
obtainable with CMS and 10 fb$^{-1}$ for the
maximal stop mixing scenario and by considering just the channels involving
the heavy Higgs bosons.
%The cross sections and BR for the Higgses are obtained from various
%dedicated packages~\cite{sigbr}.
%The coverage of the parameter space  obtainable with CMS and 10
%fb$^{-1}$ is reported in Fig.~\ref{h_discrange2} for the
%maximal stop mixing scenario and by considering just the channels involving
%the heavy Higgses.
%The cross sections and BR for the Higgses are obtained from various
%dedicated packages~\cite{sigbr} and the efficiencies of the on-line selection
%and reconstruction are taken from detailed simulation studies. 
%One can easily see that a sizable region at high $\tan{\beta}$ 
%and moderately low values of $m_A$ 
%will be covered with just 10
%fb$^{-1}$.
%Also the heavy neutral Higgs decay into a pair of muons is of
%interest thanks to the presence of the   
%five Higgs particles are predicted: one scalar light Higgs
%($h$), one scalar and one pseudoscalar two char

\section{Conclusions}
Examples of the physics reach of the CMS detector with an integrated
luminosity of 10 fb$^{-1}$ have been briefly summarised. 
Within a year about one million {\rm t} quark pairs and over 
100 million {\rm W} and {\rm Z} bosons are expected to be produced
with an electron or muon in 
the final state, which will also  allow the detector to be accurately
aligned and calibrated.
%Detailed simulation of the entire
%on-line selection chain leads to impressing numbers for the yield of
%${\rm W}$, ${\rm Z}$ and t quark when these are selected by looking at
%the presence of leptons in the final state. The numbers corresponding
%to the selection based on muons have been reported. \\
Less than 10 fb$^{-1}$ is needed for a $5\sigma$ discovery of the 
SM Higgs boson over the whole mass range from
130 to 700\,GeV/$c^2$. For smaller mass an entire year and a well
aligned and calibrated detector are needed. 
%For such low values of $m_H$ the presence of 
%forward jets deriving from the vector boson fusion
%diagram could facilitate an early discovery. 
%At higher mass values the leptonic decays into vector boson pairs
%provide significance exceeding 5$\sigma$.
A significant fraction of the MSSM parameter space is expected to be covered  
with the first 10 fb$^{-1}$ by looking for the heavy neutral Higgs
bosons H and A decays in the $\tau^+ \tau^-$ final state. 
%The favored parameter space region
%is the one at high $\tan{\beta}$ thanks to the more frequent presence
%of pairs of $b$ jets in the Higgs events.

\end{document}